# A Growth Model for Multicellular Tumor Spheroids

Pier Paolo Delsanto [†,§], Caterina Guiot [*,†], Piero Giorgio Degiorgis [†],
Carlos A. Condat [¯,ª], Yuri Mansury [#], and Thomas S. Deisboeck [#,**]

*Affiliations:* [§] Dip Fisica, Politecnico di Torino, Italy; [*]Dip. Neuroscience, Università di Torino, Italy and [†] INFM, sezioni di Torino Università e Politecnico, Italy; [¯]Department of Physics, University of Puerto Rico, Mayagüez, PR 00681, USA; [ª]CONICET and FaMAF, Universidad Nacional de Córdoba, 5000-Córdoba, ARGENTINA; [#] Complex Biosystems Modeling Laboratory, Harvard-MIT (HST) Athinoula A. Martinos Center for Biomedical Imaging, HST-Biomedical Engineering Center, Massachusetts Institute of Technology, Cambridge, MA 02139 and [**] Molecular Neuro-Oncology Laboratory, Harvard Medical School, Massachusetts General Hospital, Charlestown, MA 02129, USA.

**Corresponding author:**

Caterina Guiot
Dip. Neuroscience
Università di Torino
Corso Raffaello 30
10125 Torino
ITALY

tel: +39.11. 670.7710
fax: +39.11. 670.7708
e-mail: caterina.guiot@unito.it

## ABSTRACT

Most organisms grow according to simple laws, which in principle can be derived from energy conservation and scaling arguments, critically dependent on the relation between the metabolic rate *B* of energy flow and the organism mass *m*. Although this relation is generally recognized to be of the form *B(m)* = $m^p$, the specific value of the exponent *p* is the object of an ongoing debate, with many mechanisms being postulated to support different predictions. We propose that multicellular tumor spheroids provide an ideal experimental model system for testing these allometric growth theories, especially under controlled conditions of malnourishment and applied mechanical stress.

## 1. INTRODUCTION

As early as 1948, A.K. Solomon proposed the use of physical techniques for the development of mathematical models of both cancer growth and related therapy [1]. However, only recently have the combined power of detailed biomedical observations and increased computational capabilities enabled realistic numerical simulations [2-10]. Since growth mechanisms stem, ultimately, from cellular interactions, the above mentioned "microscopic" models should, in principle, suffice. Yet, in some cases, remarkable collective patterns are observed, which suggest the adoption of a "macroscopic" approach, not as an alternative, but as a natural complement of the microscopic models. It was recently proposed [11] that tumors may follow the general model of ontogenetic growth for all living organisms (from mammals to mollusks to plants) developed by West, Brown and Enquist (WBE) [12]. Due to its implications for tumor metastasis, cell turnover rates, angiogenesis, and invasion, the proposal of ref. 11 has immediately contributed to an ongoing debate [13].

The beauty of the WBE model is that it is entirely based on energy conservation and other general physical arguments, i.e., scaling. They start with the assumption that the energy intake, supplied by ingested nutrients, is spent partly to support the metabolic functions of the organism's existing cells and partly for cell replication, i.e. to reproduce new cells. Based on this key assumption, a universal growth curve is derived, which the authors conjectured to be applicable to all living organisms. Their conjecture is supported by data encompassing many different species [12]. Although they assumed $B \propto m^p$ with p=3/4, their value of p is not universally accepted. Indeed there is strong evidence that growth may be consistent with a *2/3* law in the case of birds and small mammals [14]. Also, in a recent study, Makarieva, Gorshkov, and Li suggest that it is the process of energy consumption and not the transport of nutrients on fractal-like networks that determines the dependence of plant energetics on size [15].

Many different explanations have been put forward for the scaling laws, ranging from four-dimensional biology [16] and quantum statistics [17] to long-bone allometry combined with muscular development [18], but, as yet, the debate is far from being settled. In this Letter we suggest that multicellular tumor spheroids (MTS) are excellent experimental model systems to test the validity of the proposed mechanisms. MTS are spherical aggregations of malignant cells [19,20], which can be grown *in vitro* under strictly controlled nutritional and mechanical conditions. Although cells appear to preserve the evolutionary self-organization rules governing the growth of the original tumor line, they also evolve according to the environmental conditions of the particular experimental setting. Here we show that MTS indeed grow following a scaling law and obtain their expected properties under conditions of malnourishment and rising mechanical stress – conditions that often apply also to tumors *in vivo*.

Before specializing to the MTS problem, we first generalize WBE's theory to arbitrary values of the scaling exponent $p$ (with the restriction that $p <= 1$). Following WBE, we assume, for simplicity, that metabolism and growth are the same for all the cells of a given individual, belonging to any given taxon, and constant throughout its lifespan. Let $B$ be the resting metabolic energy expenditure of an organism. Then, at any time $t$ and for any discrete short time interval $\Delta t$, we write the law of energy conservation as:

$$B\Delta t = Nb\Delta t + e\Delta N, \qquad (1)$$

where $N$ is the total number of cells in the organism, $b$ is the metabolic rate for a single cell, and $e$ is the energy required to create a new cell. Let $\mathbf{m}$ be the mass of a single cell and $m = N\mathbf{m}$ the mass of the whole organism. Writing $B = B_0 m^p$, Eq. (1) becomes,

$$\frac{dm}{dt} = am^p - bm, \qquad (2)$$

where $a = \mathbf{m}B_0/e$ and $b = \mathbf{b}/e$. This general equation covers the whole gamut of observed allometric power laws. For instance, $p = 3/4$ corresponds to the fractal-like distribution network of the WBE model, while $p = 2/3$ follow from the assumption that what is relevant is the nutrient entry through a two-dimensional surface that encloses a three dimensional volume. From Eq. 2 it follows that, starting from a mass $m_0$ at birth, $m$ increases (with a decreasing rate) up to a maximum body size $M = (a/b)^{1/(1-p)}$, corresponding to the zero growth point $dm/dt = 0$. Introducing the function

$$y(t) = 1 - \left(\frac{m(t)}{M}\right)^{1-p}, \qquad (3)$$

Eq. (2) becomes

$$\frac{dy}{dt} = -\mathbf{a}y, \qquad (4)$$

with $\mathbf{a} = b(1-p)$. If we now define [12] the dimensionless variable $\mathbf{t} = (1-p)bt - \ln[1-(m_0/M)^{1-p}]$, Eq. 4 has the solution,

$$y(\mathbf{t}) = e^{-\mathbf{t}}. \qquad (5)$$

which does not depend on the choice of $p$ and is therefore from the theoretical point of view universally valid.

Next we discuss some implications of these results for the specific situation of multicellular tumor spheroids, MTS. Likely, effective nutrient absorption cannot start immediately after spheroid implantation into the experimental surrounding, rather it should begin with a delay that depends on the chosen malignant cell line and on the matrix properties. We model this *start-up phase* or "accommodation time" by replacing the parameter $a$ in Eq. (2) by the time-dependent function,

$$a_1(t) = a(1 - e^{-t/T}), \qquad (6)$$

where $T$ is the *effective* accommodation time.

### A. The impact of nutrients replenishment on MTS growth

Data from Freyer and Sutherland [21] show that when the *nutrient content* of the tissue culture medium is depleted (in particular of glucose and oxygen), the growth curve of the spheroids is modified, i.e. it reaches a mass *lower* than the asymptotic value $M$. The condition of insufficient nutrient availability can be reproduced in our model by the subtraction of a term in Eq. (2), representing the missing power. It is plausible to write such a term as $f \cdot am^p$, i.e. to assume that it can be scaled as the supply term $m^p$, with the fraction $f \in [0,1]$ and $f = 0$ in conditions of full nutrient availability. Thus, assuming that the relation $\frac{dm}{dt} = (1-f)am^p - bm$ and Eq. (6) hold at any time $t$ except in equilibrium, Eq. (2) then becomes:

$$\frac{dy}{dt} + ay = af + a(1-f)e^{-t/T}. \qquad (7)$$

Equation (7) has a simple exact analytical solution whose asymptotic value $y(\infty) = f$ corresponds to the asymptotic mass

$$m_\infty = M(1-f)^{1/(1-p)} \qquad (0 \le f < 1) \qquad (8)$$

Figure (1) shows experimental data from Ref. 21 for different degrees of undernourishment. Since the available data do not allow us to select an optimal value for p, we choose p=2/3 as our working assumption for the fits (the same applies to Fig. 2). The solid lines are model fits. The agreement is excellent, except for the case of most severe undernourishment (*f = 0.8)*. This is not surprising, because a strong nutrient deficiency should also result in a substantial reduction of the metabolic rate.

### B. Impact of mechanical stress on MTS growth

Furthermore, it has been suggested that tumor growth and metastasis may be influenced by *mechanical stress*. In particular, Helmlinger et al. have investigated the volumetric growth of MTS cultured in agarose gels, showing that growth is inhibited when the gel concentration is increased [22]. This result can be explained with increasing local stress, which in turn leads to a higher cell density $r$. In fact, while $r$ is normally assumed to be constant throughout the entire growth process, the data from Ref. 22 show that the cell density may increase as a result of increasing mechanical stress within the microenvironment. Correspondingly, Freyer and Schor report that cells in the internal MTS layers experience a size reduction by up to a factor of three [23]. In order to model the impact of increasing mechanical stress, we assume that MTS exhibit some ability to be *compacted*, if not compressed, under mechanical stress, so that the cell density $r$ varies over time. As a consequence, the variable of interest is no longer the mass $m$, but the volume $v = m/r(t)$, which is the measurable quantity in real experiments. Let *V* and *R* be the asymptotic values of the volume *v* and density *r* in conditions of full nutrient availability and define, in analogy with Eq. (3),

$$z(t) = 1 - \left(\frac{v(t)}{V}\right)^{1-p}, \qquad (9)$$

With this definition, Eq. (7) becomes,

$$\frac{dz}{dt} + \left(a + \frac{d\ln g}{dt}\right)(z-1) = -\frac{a(1-f)}{g}\left(1 - e^{-t/T}\right), \qquad (10)$$

where $g(t) = [r(t)/R]^{1-p}$. The parameter *f* represents, as in Eq. (7), the coefficient of "missing power". It can also be used to model the conditions of reduced mass growth that may result as an additional effect of large mechanical stress. Equation (10) allows us to plot the volume growth curves, once the dependence of *r* on *t* is specified. To find its time dependence we note that: (*i*) both, nutrient availability and tumor growth are tightly correlated, (*ii*) increases in confining mechanical stress are a direct result of volumetric tumor growth, and (*iii*) such raising mechanical stress has a similar reducing effect on tumor growth as malnourishment, and that *r(t)* grows exponentially with the same time constant as $a_1(t)$. We then propose the following ansatz,

$$\tilde{r}(t) = R + \left(\tilde{r}_0 - R\right)e^{-t/T}, \qquad (11)$$

where $r_0$ is the initial value of the density, corresponding to the initial mass $m_o$. We expect the asymptotic value for the density, *R*, to be a monotonically increasing function of the stress, with $R = r_0$ for the unstressed spheroid. The asymptotic volume is given by,

$$v_\infty = V(1-f)^{1/(1-p)} \qquad (0 \le f < 1) \qquad (12)$$

Growth data for various gel concentrations from Ref. 22 are shown together with model fits in Fig. 2. Note the substantial increase in the accommodation time as the gel concentration (and thus the mechanical stress) is increased. These results suggest that it takes a relatively long time for the tumor cells to develop a suitable nutrient uptake within a stressed matrix. Finally, the fitting results show that the effective under-nourishment *f* is also increased with the gel concentration, while the terminal tumor volume is sharply reduced.

## 2. CONCLUSIONS

In here, we suggest that multicellular tumor spheroids can be used as an experimental model system to test allometric growth laws under controlled environmental conditions. Specifically, we have presented a theoretical model that describes their evolution under conditions of insufficient nutrient supply and/or of rising mechanical stress, and whose solutions agree with experimental data. The exponent in Eq. (8) indicates that growth is less affected by malnourishment when *p* is closer to unity. Although we derived Eq. (7) for multicellular spheroids, one could speculate that the final size of tumors or more general, organisms that grow according to the *3/4* law will be less severely modified by nutrient scarcity than the final size of those following the *2/3* law. In two-dimensional *in vitro* diffusion-controlled growth, *p = 2/3* should be replaced by *p = 1/2*, while in theory for a one-dimensional line of cells, i.e. growing in a tube and fed through its ends, we should take *p = 0*. This indicates that sensitivity to nutrient deprivation is increased in lower-dimensionality systems, such as in tumor cell aggregates growing on flat substrates in cell culture or, more importantly, invasive tumor cells forming chain-like patterns along preferred anatomically pathways, which in turn should impact their chemotactic behavior. Finally, our results for growth under stress indicate that normal nutrient uptake patterns are severely impaired by increasing mechanical stress. This should have implications for cancer growth in the neighborhood of robust tissues, such as bone.

Our conclusions should stimulate experimentalists to conduct careful 'long-term' studies of MTS growth and expansion within distinctively different microenvironmental settings. Such microtumor growth studies should then shed light not only on various important aspects of the dynamics of cancer growth, but also on the current allometric growth law controversy.

**FIGURE CAPTIONS**

**Fig. 1**: Time variation of the *under*nourished spheroid mass (see eq (3), data from Ref. 21). Solid curves are model fits for a choice of p=2/3. The y-intercept corresponds to $m_0 = 2 \times 10^{-6}$ g. Final masses $m_\yen$ are, starting from the lowest curve: *4.4 mg*, *3.7 mg*, *1.95 mg*, and *3.56×10$^{-5}$ g*. The accommodation time is, in all cases, *T = 10 h*. The values of the parameter *f* (fitting results) are given next to the corresponding curves.

**Fig. 2**: Time variation of the spheroid volume growing under different mechanical stress conditions due to various gel concentrations (see eqs: (9, 10), data from Ref. 22). Solid curves are model fits. The z-intercept corresponds to $m_0 = 4 \times 10^{-9}$ g. Final volumes and accommodation times *(V, T)* are, starting from the lowest curve: *(6×10$^{-4}$ cm$^3$, 30 h)*, *(3.8×10$^{-5}$ cm$^3$, 100 h)*, *(2.65×10$^{-5}$ cm$^3$, T = 110 h)*, *(4.88×10$^{-6}$ cm$^3$, 120 h)*. The values of *f and R* (fitting results) are given next to the corresponding curves, specified as *(f,R)*.

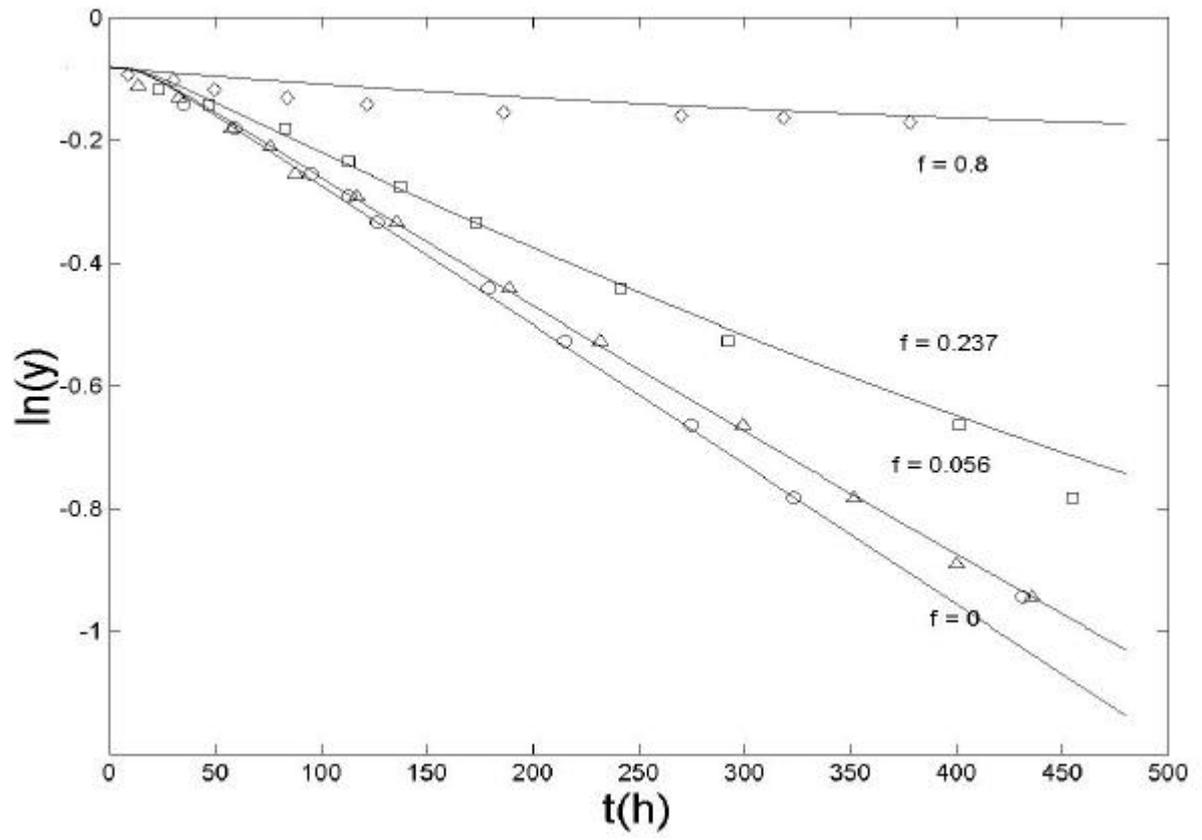

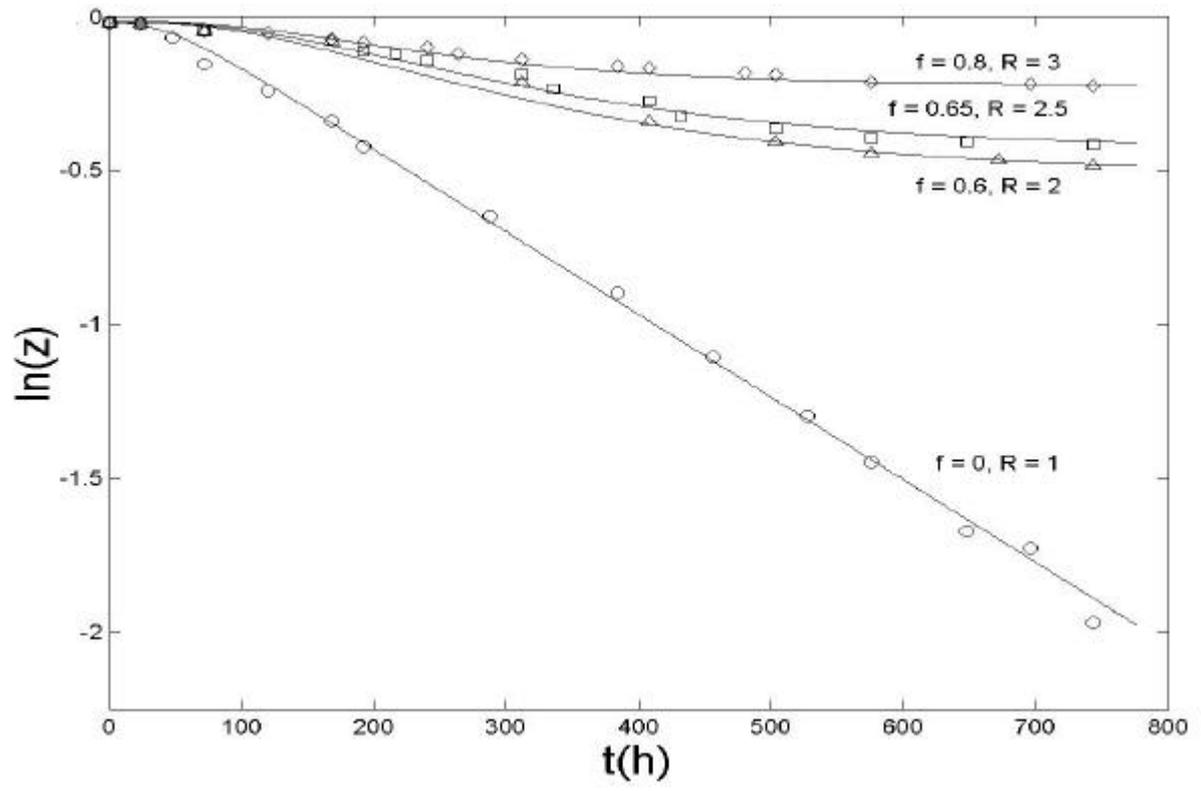